\documentclass{article}
\newcommand{\be}{\begin{equation}}
\newcommand{\ee}{\end{equation}}

\def\bea{\begin{eqnarray}}
\def\eea{\end{eqnarray}}
\def\beas{\begin{eqnarray*}}
\def\eeas{\end{eqnarray*}}
\def\simlt{\stackrel{<}{{}_\sim}}

\begin{document}

\rightline{UFIFT-HEP-06-11}
\rightline{hep-ph/0608069} \vspace*{1.5cm}

\begin{center}
{\bf \LARGE Large Lepton Mixings from Continuous Symmetries} \\
\vskip 1cm

\large{
Lisa Everett\footnote{Address after September 1, 2006: Department of Physics, University of Wisconsin, 1150 University Avenue, Madison, WI, 53706, USA.} and Pierre Ramond
\\[8mm]}

\it{Institute for Fundamental Theory, 
Department of Physics\\ 
University of Florida, Gainesville, FL, 32611, USA\\
[15mm]}
 
\large{\rm{Abstract}} \\[7mm]

\end{center}

\begin{center}

\begin{minipage}[h]{14cm}
Within the broad context of quark-lepton unification, we investigate the implications of broken continuous family symmetries which result from requiring that in the limit of exact symmetry, the Dirac mass matrices yield hierarchical masses for the quarks and charged leptons, but lead to {\it degenerate} light neutrino masses as a consequence of the seesaw mechanism, without requiring hierarchical right-handed neutrino mass terms.   Quark mixing is then naturally small and proportional to the size of the perturbation,  but lepton mixing is large as a result of degenerate perturbation theory, shifted from maximal mixing by the size of the perturbation. Within this approach, we study an illustrative two-family prototype model with an $SO(2)$ family symmetry, and discuss extensions to three-family models.  
\end{minipage}

\end{center}

\newpage 
\section{Introduction} 

Neutrino oscillation experiments have confirmed that neutrinos are massive and lepton mixing is observable.  This result, which is the most important recent experimental result in particle physics, has brought a new level of confusion and excitement to the already puzzling area of flavor physics.  Although the triumph of the Standard Model (and its extension to grand unified theories) suggests a unified
view of the quarks and leptons in terms of the gauge interactions, quarks and leptons display striking differences in terms of their masses and mixings. 

In the quark sector, the Cabibbo-Kobayashi-Maskawa mixing matrix is 
approximately the identity matrix, and the masses are equally spaced on
the logarithmic scale. Parametrized in terms of the Cabibbo angle 
$\lambda_c\equiv \sin\theta_c=0.22$ ($\theta_c\simeq 13^\circ$), the quark 
mass ratios are
\be
\frac{m_u}{m_t}\sim \lambda_c^8,\;\;\frac{m_c}{m_t}\sim 
\lambda_c^4;\;\;\;\;
\frac{m_d}{m_b}\sim \lambda_c^4,\;\;\frac{m_s}{m_b}\sim \lambda_c^2.
\ee
As a result, perturbing about the limit of vanishing Cabibbo angle is a 
popular approach to understanding the flavor problem.

In the lepton sector, the situation is quite different.   The Maki-Nakagawa-Sakata-Pontecorvo (MNSP) \cite{mnsp} lepton mixing matrix is experimentally known 
 \cite{atm,solar,chooz}
to be of the form (neglecting phases):
\be{\mathcal U}^{}_{\rm 
MNSP}~\approx~{\mathcal R}^{}_1(\theta_\oplus)\,
{\mathcal R}^{}_3(\theta_\odot),
\label{MNSP}
\ee 
in which ${\mathcal R}_a(\theta)$
denotes a rotation matrix about the $a$-axis. The atmospheric and
solar mixing angles are large:
\be
\theta^{}_\oplus~\approx~45^\circ\pm 10^\circ,\qquad  \theta^{}_\odot~\approx~34^\circ\pm 2^\circ\ .\ee 
Equally puzzling is the fact that oscillations are predominantly among two
neutrinos; the third mixing angle $\theta_{13}$ is bounded to be at 
most of Cabibbo size \cite{chooz}.
Turning now to lepton masses, the charged lepton masses display hierarchies
\be
\frac{m_e}{m_\tau}\sim \lambda_c^5,\;\;\frac{m_\mu}{m_\tau}\sim 
\lambda_c^2,
\ee
but the same is not obvious for the neutrinos.  Oscillation 
data has indicated that the ratio of the observed solar  
and atmospheric mass-squared differences 
can be expressed in terms of $\lambda_c$ as follows:
\be
\label{rdef}
\frac{\Delta m^2_\odot}{\Delta m^2_\oplus}\simeq \frac{8\times 
10^{-5}\,{\rm eV}^2}{2\times 10^{-3}\,{\rm eV}^2}\sim \lambda_c^2.
\ee
Assuming only three active 
neutrinos oscillate, with masses $m_{1,2,3}$, 
there are several possible neutrino mass patterns, all of which display milder hierarchies than those of the charged fermions:
\begin{itemize}
\item Normal hierarchy ($m_1<m_2\ll m_3$):
\be
\frac{m_1}{m_3}\sim \sqrt{\left (\frac{m_2}{m_3}\right 
)^2-\lambda_c^2},\;\;
\frac{m_2}{m_3}\sim \lambda_c, 
\ee
\item Inverted hierarchy ($m_3\ll m_1\sim m_2$): 
\be
\frac{m_1}{m_2}\sim 1+\lambda_c^2,\;\;
\frac{m_2}{m_3}\simlt \lambda_c.
\ee
\item Quasi-degenerate ($m_1\sim m_2 \sim m_3$):
\be
\frac{m_1}{m_3}\sim\frac{m_2}{m_3}\sim 
1+\lambda_c,\;\;\frac{m_1}{m_2}\sim 1+\lambda_c^3.
\ee
\end{itemize}
The data have challenged us to understand the origin of the observed discrepancies between the quark and lepton sectors.   This question is particularly intriguing within the context of quark-lepton unification \cite{Pati:1973uk}, for which all available data can be synthesized in the search for a compelling flavor theory.  The lepton data has led to a recent renaissance in flavor model building, with many examples in the literature (see the reviews \cite{nureviews} and references therein).

The first step in addressing the flavor puzzle is to seek a reasonable theoretical starting point for building flavor theories.  To this end, quark-lepton unification compels us to consider the effects of small (e.g. Cabibbo sized) perturbations in the lepton sector.  For the mixings, $\theta_{13}$ is akin to quark angles (deviation from zero), while for the solar and atmospheric mixings, small parameters would appear as {\it deviations} from large initial values. The measured angles are thus ``hazed" away from starting values which assume values indicative of flavor symmetries \cite{bimax,pierrerefs,haze} (in other approaches  $\theta^{}_\oplus$ and $\theta^{}_\odot $ take their {\it central} experimental values as indicative of discrete flavor symmetries \cite{discretesymm}).\footnote{Alternatively, one can give up the study of flavor physics, appeal to some landscape, or like number theorists of old, ascribe magical meanings to the central values of the angles (our favorite is the Phidian angle (golden mean) $\pi/5\approx 36^\circ$).}  Small parameters may govern a subset of the neutrino masses or their deviations from a common mass.   However, given the large lepton mixings, a natural theoretical starting point is neutrino mass degeneracy,  since large angles can arise from small perturbations about degenerate structures.

In that spirit, we investigate the implications of the family symmetries which emerge by {\it requiring} a theoretical starting point where in the limit of exact symmetry, the Dirac ($\Delta I_w=1/2$) and Majorana ($\Delta I_w=0$) mass matrices yield hierarchical masses for the charged fermions and degenerate masses for {\it both} the light neutrinos obtained from the seesaw mechanism \cite{SEESAW} and the heavy right-handed neutrinos.  These requirements lead to specific constraints on the $\Delta I_w=1/2$ sector which suggest features of the family symmetry.  Note that the path to neutrino mass degeneracy differs here from that most often found in the literature (e.g. in single right-handed neutrino dominance models \cite{srnd}), in which hierarchical right-handed neutrino masses offset the hierarchical Dirac masses. In its simplest implementation, an $SO(2)$ family symmetry results from these requirements. Large lepton mixing angles then arise from small perturbations in the neutrino sector, as expected from degenerate perturbation theory.  

In this paper, after arguing that quasi-degenerate neutrinos are a reasonable starting point on both phenomenological and theoretical grounds,  we outline our approach for an arbitrary number of families (assuming one right-handed neutrino per family).  We then focus in detail on a prototype two-family model with the $SO(2)$ family symmetry which emerges naturally in our approach.  In this scenario, the  charge eigenstates  are given by
$$\frac{1}{\sqrt{2}}(\psi\pm i\psi^\prime),$$
where $\psi$ and $\psi^\prime$ are the current eigenstates of the two families. The degeneracy of the light seesawed neutrinos is ultimately due to nontrivial phases in the couplings enforced by the symmetry.   With symmetry breaking in either the $\Delta I_w=0$ or the $\Delta I_w=1/2$ neutrino sector, the degeneracy is lifted and the mixing angle is hazed away from $45^\circ$ by the perturbation. We next discuss ways to extend the general approach to three-family models, and conclude by discussing  avenues of future exploration.

\section{Theoretical Framework}
\label{mainapproach}

\subsection*{The Case for Neutrino Mass Degeneracy}
In our theoretical approach to the flavor puzzle, the discrepancies between the quark and lepton mixings  are ultimately due to the theoretical requirement of neutrino mass degeneracy in the limit of an exact family symmetry.   Hence, we first wish to emphasize that quasi-degenerate neutrinos are intriguing both from a phenomenological and a theoretical point of view.  In particular, while the overall neutrino mass scale has been constrained  from cosmology \cite{cosmo} (and, to a lesser extent, $0\nu\beta\beta$ \cite{nuless}), quasi-degenerate scenarios are consistent with the data (though with a reduced parameter space).  The cosmological constraints  yields the following bound on the sum of the neutrino masses \cite{cosmo}:
\be
\sum_i m_i \simlt  0.2 - 2 \;  {\rm eV}.
\ee
The range reflects the dependence of the bound on priors and the data sets used for the fits.  Assuming a conservative estimate of the bound of $0.9$ eV, the overall neutrino mass scale $m_0$ for quasi-degenerate models is $m_0\simlt 0.3\,{\rm eV}$.   Such scenarios have the advantage that they will be probed by neutrinoless double beta decay experiments \cite{nuless} and direct beta decay experiments \cite{Osipowicz:2001sq}.

From a theoretical standpoint, one may worry that if the experimental bound $m_0$ is not much larger than the observed mass splittings,  the ratio of the mass-splittings to the overall scale is not a reliable expansion parameter.  This observation, which would naively appear to disfavor degenerate neutrinos as a theoretical starting point,  can be countered in two ways.  First, note that current bounds do not preclude scenarios in which this ratio is a small parameter.  To see this more clearly, 
let us parametrize the masses as follows:
\be
m_1=m_0-|\Delta_\odot|,\;\;m_2=m_0,\;\;m_3=m_0+\Delta_\oplus,
\ee  
in which $\Delta_\oplus$ can take either sign.  The experimentally measured mass-squared differences are then given by
\bea
\Delta m_\oplus^2&\equiv&|m^2_3-m^2_2|=(2m_0+\Delta_\oplus)|\Delta_\oplus|\nonumber \\
\Delta m_\odot^2&\equiv&|m^2_2-m^2_1|=(2m_0-|\Delta_\odot|)|\Delta_\odot|.
\eea
The ratios of the mass-splittings to the overall mass scale take 
the form
\bea
\frac{\Delta_\oplus}{m_0}&=&\pm \sqrt{1\pm \frac{\Delta 
m^2_\oplus}{m_0^2}}-1\nonumber \\ 
\frac{|\Delta_\odot |}{m_0}&=&\pm \sqrt{1-\frac{\Delta 
m^2_\odot}{m_0^2}}+1.
\eea
These ratios can be large numbers if $m_0 \ll \sqrt{\Delta 
m^2_{\rm (exp)}}$; however, this only occurs when $m_0$ is more than an order of magnitude below the cosmological limit.   (The solutions with $\mathcal{O}(1)$ ratios independently of $m_0$ correspond to negative neutrino masses when $\Delta m^2_{{\rm (exp)}}\rightarrow 0$, and hence effectively cover the same regime.)  

In addition, note that $\mathcal{O}(1)$ or larger ratios are not necessarily problematic for seesaw models based on small perturbations about degenerate structures. Since the seesaw is a convolution of the $\Delta I_w=1/2$  and  $\Delta I_w=0$  mass matrices, we will see that small perturbations in the Majorana masses can be enhanced by the hierarchies in the Dirac mass terms without invalidating perturbation theory.  In the two-family $SO(2)$ model discussed later in the paper, the Dirac masses are $m_{1,2}\sim e^{\pm \eta}$  and there is a small parameter $\delta \ll 1$ in the electroweak singlet sector, yet the mass-squared difference is $\sim \delta \cosh 2\eta $ and the shift in the mixing angle is $\sim \delta \sinh 2\eta$.  Therefore, $\Delta_\oplus/m_0$ and $\Delta_\odot/m_0$ need not be small even when there are small parameters in the seesaw matrix.   This is a point that, to our knowledge, has not been fully appreciated in the literature.

\subsection*{Systematics of the Approach}
With these ideas in mind, we now outline our theoretical approach.  The main idea is to study the symmetries which emerge upon imposing the theoretical requirement that at tree level\footnote{We focus here on tree-level predictions.  In complete scenarios one would need to consider higher-dimensional operators a la Froggatt-Nielsen \cite{fn}; we relegate this for future study.}  the charged fermion masses are hierarchical, but the light neutrino masses emerge naturally through the seesaw mechanism, without strong hierarchies in the right-handed neutrino sector.\footnote{It is worth noting, however, that some hierarchy in the right-handed neutrino sector may be advantageous for leptogenesis \cite{leptogenesis} (but see, however, \cite{pilaftsis}).} These requirements lead to basis-dependent conditions on both the $\Delta I_w=0$ and the $\Delta I_w=1/2$ mass matrices, which can be expressed in the  basis in which the charged currents are diagonal (the current basis) as follows:\\
 
 \noindent $\bullet$ {\bf $\Delta I_w=0$ sector}. 
 The Majorana mass matrix for the right-handed neutrinos is proportional to the identity matrix:
 \be
 \label{majcond}
 \mathcal{M}_{\rm Maj}\propto 1.
 \ee
Assuming one right-handed neutrino per family, the symmetry is $O(n)$ (for $n$ families).  If the number of right-handed neutrinos per family is doubled, as in $E_6$ models \cite{Gursey:1975ki}, the symmetry can be larger, e.g. $SU(n)$. \\ 
 
 \noindent $\bullet$ {\bf $\Delta I_w=1/2$ sector}. 
 Hierarchical Dirac mass eigenvalues require that the Dirac mass matrices $\mathcal{M}_D$ obey the following condition:
 \be
 \mathcal{M}_D\mathcal{M}_D^\dagger\neq 1.
\ee 
We require that this condition holds for the Dirac mass matrices for both the charged and neutral fermions.  Given the form of  $\mathcal{M}_{\rm Maj}$ above and recalling the form of the seesaw matrix,
\be
\mathcal{M}_\nu=\mathcal{M}_D\mathcal{M}_{\rm Maj}^{-1}\mathcal{M}_D^T,
\ee
degenerate light neutrino masses require an additional condition\footnote{This condition holds because the observable quantities are the absolute values of the mass eigenvalues, and encompasses the special case in which $ \mathcal{M}_D\mathcal{M}_D^T\propto 1$.}
 on the neutral Dirac mass matrix (which may also hold for the charged fermions):
 \be
 \label{degcond}
(\mathcal{M}_D\mathcal{M}_D^T)(\mathcal{M}_D\mathcal{M}_D^T)^\dagger \propto 1.
\ee
Restricting ourselves for now to scenarios with an equal number of right-handed and left-handed neutrinos, this condition can be reexpressed as in terms of the Hermitian combination $\mathcal{H}\equiv \mathcal{M}_D^\dagger \mathcal{M}_D$ as follows:
\be
\label{hcond}
\mathcal{H}\mathcal{H}^T\propto 1.
\ee
Writing $\mathcal{H}=\mathcal{S}+i\mathcal{A}$, in which $\mathcal{S}$ and $\mathcal{A}$ are real symmetric and antisymmetric matrices, respectively, Eq.~(\ref{hcond}) translates to the conditions

$$ \mathcal{S}^2+\mathcal{A}^2 \propto 1;\;\;\;  [\mathcal{A},\mathcal{S}]=0,$$
which also imply $$[\mathcal{A},\mathcal{H}]=0.$$
$\mathcal{A}$ is a real antisymmetric matrix and thus is in the Lie algebra of $O(n)$.  Hence,  $\mathcal{H}$ commutes with at least one generator of $O(n)$, requiring an $SO(2)$ symmetry at least as part of the family symmetry of the theory.   Eq.~(\ref{hcond}) also has implications for the Dirac mass hierarchy; the eigenvalues of $H$ are either (up to an overall scale) 1 or mutually reciprocal pairs $e^{\pm \eta}$.  For odd $n$, the middle Dirac mass eigenvalue is the geometric mean of the highest and lowest eigenvalues.\footnote{Note that this also holds for models based on rational hierarchy \cite{kausmeshkov}.}
 
The above discussion did not specify the number of families. Beginning with the case of two families for simplicity, we note that the unique solution for $\mathcal{H}$ is  
\be
\mathcal{A}=c_A i\sigma_2,\;\;\; \mathcal{S}= c_S 1,
\ee 
where $c_{A,S}$ are arbitrary real constants.  Writing $\mathcal{M}_D$ as a sum of group elements $T_i$ with complex coefficients $\omega_i$, such that 
 \be
 \mathcal{M}_D=\omega_1T_1+\omega_2 T_2,
 \ee  
 one solution is $T_1=1$ and $T_2=i\sigma_2$, with $c_S=\omega_1^2+\omega_2^2$, and $c_A=-2 {\rm Im}( \omega_1\omega_2^*) $.\footnote{This result not only satisfies Eq.~(\ref{degcond}), but the stronger condition that $\mathcal{M}_D\mathcal{M}_D^T\propto1$.}   At least one complex $\omega_i$ is required, or the hierarchical structure of $\mathcal{M}_D$ is lost (the eigenvalues form a mutually reciprocal pair).  We will  present explicit two-family $SO(2)$ models with  these properties in Section~\ref{twofamsect}.
 
 For three families, a generalization to $O(3)$ is suggested by the requirement on the $\Delta I_w=0$ mass terms as given in Eq.~(\ref{majcond}).  However, the requirement that $\mathcal{H}$ commutes with all of the generators of $O(3)$ implies that $\mathcal{H}$ must be proportional to the identity, and hence there is no hierarchy in the Dirac mass eigenvalues.  Therefore, an $O(3)$ symmetry in the $\Delta I_w=1/2$ sector does not satisfy the constraints of our approach.  For another way to see this, 
 let us write $\mathcal{M}_D$ once again as a sum of group elements, 
 \be
 \mathcal{M}_D=\omega_1T_1+\omega_2T_2+\omega_3T_3,
 \ee
and require that $\mathcal{M}_D\mathcal{M}_D^T\propto 1$. For $O(n)$, $T_iT_i^T=1$,  leading to the conditions
 \be
 \omega_1\omega_2(T_1T_2^T+T_2T_1^T)+\omega_1\omega_3(T_1T_3^T+T_3T_1^T)+\omega_2\omega_3(T_2T_3^T+T_3T_2^T)=0.
 \ee
For $O(3)$, these conditions cannot be satisfied by inspection.  One $T_k$ must be chosen such that $T_i T_k^T=0$ for $i\neq k$, corresponding to an $O(2)$ subgroup.  

 For $n>3$, in principle it is possible to find at least one other generator of $O(n)$ which commutes with $\mathcal{A}$ and some $\mathcal{S}$'s. 
We will continue the discussion of how to generalize the approach to three and more families in Section~\ref{threefamsect}.

\section{Two Family Models}
\label{twofamsect}
We now explore prototype two-family models which satisfy the theoretical guidelines outlined above.  The main assumption is the presence of an $SO(2)$ family symmetry which acts on the following linear combinations of the current eigenstates\footnote{All fermions are taken to be left-handed. Barred fields denote the right-handed states.} $\psi$ and  $\psi'$ (which have the same Standard Model quantum numbers)
\be
\psi^{}_\pm~\equiv~ \frac{1}{\sqrt{2}}(\psi\pm i\psi_{}^\prime)\ ,
\ee
according to
\be
\psi_\pm\,\rightarrow\,e^{\pm i\,\alpha}_{}\,\psi_\pm.
\ee
We avoid labeling the current eigenstates numerically to avoid confusion with the mass eigenstates $\psi^{(m)}$, which are related to the current eigenstates $\psi^{(c)}$ by
\begin{eqnarray}
\psi^{(c)}\equiv~  \left(\begin{array}{c }
\psi
\\
\psi^\prime
\end{array}\right)=\mathcal{U}\psi^{(m)}~\equiv \mathcal{U}\left(\begin{array}{c }
\psi^{}_1
\\
\psi^{}_2
\end{array}\right)\ .
\end{eqnarray}
In the current basis, the Dirac mass matrices for fermions of charge $q$  are
\be
\mathcal{M}^{(q)}=\mathcal{U}^T_q\mathcal{D}^{}_q\mathcal{V}^{}_q,
\ee
with left- and right-handed mixings $\mathcal{U}_q$ and $\mathcal{V}_q$, respectively.  $\mathcal{D}_q$  is the diagonal matrix of mass eigenvalues.  The seesaw matrix $\mathcal{M}_\nu$ for the light neutrinos  is
\be
\mathcal{M}^{}_\nu~=~\mathcal{U}_\nu^T\mathcal{D}^{}_\nu\mathcal{U}^{}_\nu,
\ee
and hence in our conventions, the lepton mixing matrix is 
$\mathcal{U}_{\rm MNSP}=\mathcal{U}_{-1}\mathcal{U}_\nu^\dagger$.

\subsection*{$SO(2)$ Symmetry Limit}
Let us begin with the neutral lepton sector.
The $SO(2)$ symmetry operates on the lepton 
doublets $L_\pm$ and the right-handed neutrinos $\overline{N}_\pm$, which then enforces degenerate right-handed neutrino masses: 
$$2\,M\,\overline N^{}_+\,\overline 
N^{}_-~=~M\,(\overline{N}\,\overline{N}+\overline{N}_{}^\prime \overline{N}_{}^\prime)\ ,$$
so that the Majorana mass matrix ${\mathcal M}^{}_{\rm Maj}$ is proportional to the identity matrix.
The $SO(2)$-invariant neutral Dirac mass terms are
$$a\,\nu^{}_+\,{\overline N}^{}_-+b\,\nu^{}_-\,{\overline N}^{}_+ \ ,$$
such that in the current basis the neutral Dirac mass matrix $\mathcal{M}_D^{(0)}$ is 

\be
\mathcal{M}^{(0)}_D=\pmatrix{z&-iw\cr iw&z}\ ,
\ee
where $z=(a+b)/2$, and $w=(a-b)/2$.  Note that this matrix is of the form
\be
\mathcal{M}^{(0)}_D=z 1-i w (i\sigma_2),
\ee
which satisfies the conditions given in Eqs.~(\ref{degcond})--(\ref{hcond}).
It has the properties that its eigenvalues are nondegenerate  
\be
m^{D}_+~=~|z+w|~=~|a|,\;\;m^{D}_-~=~|z-w|~=~|b|,
\ee
and the left-handed fields are maximally mixed: 
\be
\mathcal{U}^{}_{0}~=~\frac{1}{\sqrt{2}}\pmatrix{1&-i\cr 
1&i}=\mathcal{R}(-\pi/4)\mathcal{P}\ ,
\ee
in which $\mathcal{R}(\theta)$ denotes a ($2\times 2$) rotation by $\theta$ and 
$\mathcal P$ is a diagonal phase matrix:
\be
\mathcal{P}~=~\pmatrix{1&0\cr 0&i}.
\ee
However, the {\em physical} neutrino masses $m_{\nu_{1,2}}$ are degenerate:
\be
\mathcal{M}_\nu~=~\mathcal{M}_D^{(0)} \mathcal{M}_{\rm Maj}^{-1} 
(\mathcal{M}_D^{(0)})^T~=~\frac{1}{M}\,\pmatrix{z^2-w^2&0\cr 0&z^2-w^2},
\ee
and there is no observable mixing.  The $SO(2)$ symmetry requires both the 
right-handed and the left-handed (through the seesaw) neutrino masses to be fully degenerate, even though the Dirac masses can be hierarchical (one can even vanish if $z=\pm w $).  
For real  $z$ and $w$, a useful parametrization is
$$
z~=~\mu\,\cosh\eta\ ,\qquad 
w~=~\mu\,\sinh\eta\ ,$$
where $\mu$ is a mass parameter; alternatively
$$
a~=~\mu\, e^{\,\eta}_{}\ ,\qquad b~=~\mu\, e^{-\eta}_{}\ ,$$
suggestive of Yukawa couplings proportional to warp factors.

Turning to the charged lepton sector, it is simplest to assume that under the family symmetry, the right-handed charged leptons transform as 

\be
\overline e^{}_\pm~\rightarrow~ e^{\pm i\alpha}_{}\overline e^{}_\pm .
\ee
$SO(2)$ invariance yields the  charged lepton Yukawa couplings
$$f^{}_e\,e^{}_+\overline e^{}_-+h^{}_e\,e^{}_-\overline e^{}_+,
$$
so that in the current basis, the charged lepton Dirac mass matrix is 
 \be
{\cal M}_D^{(-1)}~=~\pmatrix{u&-i\,v\cr i\,v& u}\ ,
\ee
where $u=(f_e+h_e)/2$ and $v=(f_e-h_e)/2$. The analysis proceeds in a 
similar way to that of the neutral Dirac masses.  The  mass eigenvalues are given by 
\be
m^{}_{e_1}~=~\vert\,u+v\,\vert=\vert \,f_e \,\vert ,\qquad 
m^{}_{e_2}~=~\vert\,u-v\,\vert=\vert \,h_e \,\vert 
\ .
\ee
If the Yukawa couplings are real we can use the same parametrization, and 
set 
\be
f^{}_e~=~\mu^{}e^{-\eta_e}\ ,\qquad h^{}_e~=~\mu^{}\,e^{\,\eta_e}\ .\ee
The diagonalization produces the same type of left-handed mixing matrix, {\it i.e.}
\be
{\mathcal 
U}_{-1}^{}~=~\mathcal{R}(-\pi/4)\mathcal{P}\ .\ee
Even though the charged lepton mass matrix is diagonalized by a maximal 
$45^\circ$ rotation, lepton mixing is unobservable because of the neutrino mass degeneracy.

If the neutrinos had only Dirac mass terms, the large angle rotation of the charged lepton sector would be completely undone by the large angle rotation of the neutrino sector in the symmetry limit.  Symmetry breaking in the $\Delta I_w=1/2$ sector would yield a nonvanishing mixing angle, but since this sector is non-degenerate, the mixing would be proportional to the symmetry breaking parameter and therefore small, in contradiction with the neutrino data.  The seesaw mechanism is therefore crucial in leading to the possibility of large observable mixing from small symmetry breaking effects.

Finally we note that we could have considered different family charge 
assignments for the right-handed charged leptons. There is a special case 
where ${\overline e}$ and ${\overline e}^\prime$ have the 
same $SO(2)$ charge:

\be
\overline{e}~\rightarrow~ e_{}^{-i\,\alpha}\,{\overline 
e}\ ;\qquad  {\overline e}^{\prime}~ \rightarrow~ 
e^{-i\,\alpha}\,{\overline e}^{\prime},
\ee
{\it i.e.}, opposite of $L_+$. The allowed charged lepton mass terms are 
$$
e^{}_+\,(c\,{\overline e}+d\,{\overline 
e}^{\prime}\,)\ .
$$
The coupling to the other combination $e_-$ is forbidden by the 
symmetry, or relatively suppressed, as it would be in the 
Froggatt-Nielsen approach. The charged lepton mass 
matrix is now of a different form:
\be
{\cal M}_D^{(-1)}=\frac{1}{\sqrt{2}}\pmatrix{c&d\cr i\,c& i\,d}\ .
\ee
The charged fermion masses are

$$ m^{}_{e_1}~=~0\ ,\qquad 
m^{}_{e_2}~=~\sqrt{c\,{\overline c}+d\,{\overline d}}\ .$$ 
The left-handed mixings are again given by
\be
{\mathcal 
U}_{-1}^{}~=~
\mathcal{R}(-\pi/4)\mathcal{P}\ .\ee
Again, athough the charged lepton mass matrix is diagonalized by a 
maximal $45^\circ$ rotation, it yields no observable mixing due to the neutrino mass degeneracy. 

This charge assignment does not lead to qualitatively different physics from the 
previous case, except that one of the charged leptons is massless. However, it 
is not as natural as the previous charge assignment if we assign the same $SO(2)$ charges 
to complete grand-unified families.

We assume that the family symmetry acts in the same way on the quarks,

\be 
{\bf u}^{}_\pm~\rightarrow~e^{\pm i\alpha}_{}{\bf u}^{}_\pm\ ,\qquad {\bf d}^{}_\pm~\rightarrow~e^{\pm i\alpha}_{}{\bf d}^{}_\pm\ ,\ee
leading to the mass matrices: 
$$
(f_u\,{\bf u}^{}_+\overline{\bf u}^{}_-+ h_u{\bf u}^{}_-\overline{\bf u}^{}_+)~+~(f_d{\bf d}^{}_+\overline {\bf d}^{}_-+ h_d{\bf d}^{}_-\overline {\bf d}^{}_+)\ .
$$
The analysis proceeds exactly as before: all four quarks are massive, 
but since the two Dirac matrices are diagonalized in the same way, the CKM mixing matrix is proportional to the identity (which is not a bad starting point).   

To summarize this model in the limit of exact symmetry, we see that all charged particles are massive. 
Quarks and charged leptons with the same quantum numbers have different 
masses while the two left-handed (and the two right-handed) neutrinos are mass degenerate. None of these particles mix;  both families are stable. Next we explore the effects of 
family symmetry breaking.  


\subsection*{Symmetry Breaking}
Observable lepton mixings require distinct neutrino masses: our $SO(2)$ 
family symmetry must be broken.  The origin of this  
breaking is not known, but in principle it can occur both in the 
$\Delta I_w=1/2$ and the $\Delta I_w=0$ sectors. \\ \\

\noindent $\bullet$ {\bf $\Delta\,I_w=0$ Sector}.
If the flavor symmetry breaking 
lies entirely in the $\Delta\,I_w=0$ mass matrix, we can expect new effects only in the neutrino 
sector; all electroweak doublet Dirac matrices remain unaffected. Since 
the Majorana matrix is symmetric, the symmetry-breaking terms can occur along  $\sigma_3$ and/or  $\sigma_1$.  

Let us begin by considering symmetry breaking along $\sigma_3$, such that\be
\mathcal{M}^{}_{\rm Maj}~=~M\,\pmatrix{1+\delta&0\cr 0& 1-\delta}\ ,\ee
where $\delta <<1$ is the symmetry-breaking parameter. The degeneracy of the 
right-handed neutrinos is broken, yielding the seesaw matrix  

\be
\mathcal{M}^{}_\nu~=~\hat\mu\, 
\pmatrix{1-\delta\cosh2\eta&-i\delta\sinh2\eta\cr 
-i\delta\sinh2\eta&1+\delta\cosh2\eta}\ ,\ee
with $\hat\mu \equiv \mu^2/(M(1-\delta^2))$. The Hermitian combination

\be
\mathcal{M}^{}_\nu \mathcal{M}_\nu^\dagger
~=~ \hat\mu^2\,\pmatrix{1-2\delta\cosh 
2\eta+\delta^2\cosh 4\eta&-i\delta^2\sinh 4\eta\cr i\delta^2\sinh 
4\eta&1+2\delta\cosh 2\eta+\delta^2\cosh 4\eta}
\ee
has distinct eigenvalues:
\be
m^2_{\nu_{1,2}}=\hat{\mu}^2\left(1+\delta^2\cosh 4\eta \mp\ 2\delta \cosh 2\eta\sqrt{1+\delta^2\sinh^2 2\eta}\,\right)\ .
\ee
The mass-squared difference is therefore of $\mathcal{O}(\delta)$:

\be \Delta m^2~=~|\,m_{\nu_2}^2-m_{\nu_1}^2\,|~=~4\delta\hat\mu^2
\cosh 2\eta\sqrt{1+\delta^2\sinh^2 2\eta}\ .
\ee
For small $\delta$, the left-handed mixing matrix takes a nearly diagonal form
\be
{\mathcal U}_\nu=\pmatrix{1&-\frac{i\delta}{2}\sinh 2\eta\cr - 
\frac{i\delta}{2}\sinh 2\eta&1}+\mathcal{O}(\delta^2)\ ,\ee
providing a small shift from the maximal mixing angle generated in the charged lepton sector:

\be
{\mathcal U}_{\rm MNSP}~=~ 
\frac{1}{\sqrt{2}}\,\pmatrix{1+ \frac{\delta}{2}\sinh 
2\eta&-(1-\frac{\delta}{2}\sinh 2\eta)\cr 1- \frac{\delta}{2}\sinh 
2\eta&1+\frac{\delta}{2}\sinh 2\eta}\mathcal{P} 
+\mathcal{O}(\delta^2).
\ee
Family symmetry breaking is seen to generate an observable MNSP angle:
\be
\theta=\frac{\pi}{4}-\frac{\delta\sinh 2 \eta}{2}+\mathcal{O}(\delta^2),
\ee 
hazed away from $45^\circ$ by a slightly broken continuous family 
symmetry.  As anticipated,  degenerate perturbation can create a large 
angle; in this case the small parameter enters as deviation from $\pi/4$.  


Let us now consider symmetry breaking  along $\sigma_1$, for which 
\be
{\mathcal M}^{}_{\rm Maj}~=~M\,\pmatrix{1&\delta\cr \delta&1}\ ,\ee
which leads to the seesaw mass matrix

\be
\mathcal{M}^{}_\nu~=~\hat\mu\,\pmatrix{1+i\delta\,\sinh2\eta&-\delta\cosh2\eta\cr 
-\delta\cosh2\eta&1-i\delta\,\sinh2\eta}\ .
\ee
The diagonal elements of the Hermitian combination $ \mathcal{M}^{}_\nu\mathcal{M}_\nu^\dagger $ are degenerate:

\be
\mathcal{M}^{}_\nu\mathcal{M}_\nu^\dagger
~=~ \hat\mu^2\,\pmatrix{1+\delta^2\cosh 4\eta&-2\delta\cosh 
2\eta(1+i\delta\sinh 2\eta)\cr-2\delta\cosh 2\eta(1-i\delta\sinh 
2\eta)&1+\delta^2\cosh 4\eta}\ ,
\ee
such that the left-handed mixings are given by
\be
\label{sigma1mix2family}
{\mathcal U}^{}_\nu~=~\mathcal{P}_\nu\frac{1}{\sqrt{2}}\,\pmatrix{1&e^{-i\rho}\cr 
-e^{i\rho}&1}\ ,
\ee
which describes a maximal mixing angle.  In the above,   

\be
\tan \rho~=~\delta\sinh 2\eta\ ,\ee
and the nontrivial phase matrix is

\be 
\mathcal{P}^{}_\nu=\pmatrix{e^{i\rho/2}&0\cr 0&e^{-i\rho/2}}\ .
\ee
The eigenvalues are the same as for diagonal breaking, and once again

\be
\Delta m^2~=~|\,m_{\nu_2}^2-m_{\nu_1}^2\,|~=~4\delta\hat\mu^2\cosh2\eta \sqrt{1+\delta^2\sinh^2 2\eta}  \ .\ee
Although $\mathcal{U}_{-1}$ and $\mathcal{U}_\nu$ contain  $45^\circ$ rotations,  
the MNSP matrix takes the form
\be
{\mathcal U}^{}_{\rm  
MNSP}~=~
\frac{1}{\sqrt{2}}{\pmatrix{\sqrt{1+\sin\rho}&-\,e^{-i\rho}\,\sqrt{1-\sin\rho}\cr 
\,e^{i\rho}\,\sqrt{1-\sin\rho}&\sqrt{1+\sin\rho}}\mathcal{P}\mathcal{P}^{*}_\nu} 
+\mathcal{O}(\delta^2)\ ,
\ee
which, for small $\delta$, describes a rotation angle $\theta$, slightly 
hazed away from $45^\circ$: 

\be
\tan\theta~=~\sqrt{\frac{1-\sin\rho}{1+\sin\rho}}\approx~ 
1-\delta\sinh 2\eta\ ,
\ee
such that as in the case of diagonal breaking,

\be
\theta~=~\frac{\pi}{4}-\frac{\delta\sinh 2 \eta}{2}+\mathcal{O}(\delta^2)\ .
\ee
It can be shown that symmetry breaking along $\sigma_1$ and $\sigma_3$ are identical up to phases in the MNSP matrix.  For off-diagonal breaking, 
nontrivial phases prevent a cancellation of the large mixings of the charged and neutral leptons,
resulting in a nearly maximal observable mixing angle. 
These phases include the observable CP-violating Majorana phase present in the two family case. 
\\ \\

\noindent $\bullet$ {\bf $\Delta\,I_{\rm w}=1/2$ Sector}.   As previously stated, quark mixings can arise only if the family symmetry is broken in the electroweak doublet sector. In principle such
breaking can be generated through radiative effects stemming from the
breaking in the $\Delta I_w=0$ sector, but for the present purposes, let us
assume that symmetry breaking effects arise solely in
the $\Delta I_w=1/2$ sector of the theory.  Such effects can occur along with the $SO(2)$ invariant Yukawa couplings as
follows:

\be
a \nu^{}_+\overline{N}^{}_-+b\nu^{}_-\overline{N}^{}_+ +\mu 
\delta\nu^{}_+\overline{N}^{}_+  
+\mu \delta \nu^{}_-\overline{N}^{}_-\ ,
\ee
where we have assumed that the breaking occurs in the $\vert \Delta 
F=2\vert$ terms only, with strength $\delta$.  In the current basis, the Dirac mass matrix now 
becomes  
\be
\mathcal{M}^{}_D~=~\mu \pmatrix{\cosh\eta+\delta&-i\sinh \eta\cr i \sinh\eta & 
\cosh\eta-\delta}\ .
\ee
The seesaw matrix then takes the form
\be
\mathcal{M}^{}_\nu~=~\mathcal{M}^{}_D\mathcal{M}_D^T~=~\frac{\mu^2}{M}\pmatrix{1+2\delta\cosh\eta 
+\delta^2&2i\delta\sinh\eta\cr 2i\delta\sinh\eta&1-2\delta\cosh\eta
+\delta^2}\ .
\ee
To leading order in $\delta$, the masses are

\be
m^2_{1,2}=\frac{\mu^2}{M}(1\mp 4\delta\cosh\eta 
+\mathcal{O}(\delta^2))\ ,
\ee
and the mixing matrix is  

\be
\mathcal{U}_\nu=\pmatrix{1&i\delta\sinh\eta 
\cr i \delta \sinh\eta &1}+\mathcal{O}(\delta^2)\ .
\ee
Hence, the mass-squared difference and the MNSP mixing angle are 
\be
\Delta m^2~=~8\delta\cosh\eta+\mathcal{O}(\delta^2)\ ,\qquad
\theta~=~\frac{\pi}{4}+\delta \sinh\eta+\mathcal{O}(\delta^2)\ .
\ee
For the neutrinos, the results are essentially the same as for 
electroweak-singlet breaking, with the seesaw as the crucial ingredient for generating large mixing. 

This can best be seen by considering the effect of this breaking in the quark sector. We  take the quark mass matrices to be           

\be
m^{}_u(\,e^{-i\eta_u}_{}\,{\bf u}^{}_+\overline{\bf u}^{}_-+e^{i\eta_u}_{}{\bf u}^{}_-\overline{\bf u}^{}_+)+m^{}_u\delta^{}_u(\,{\bf u}^{}_+\overline{\bf u}^{}_++ {\bf u}^{}_-\overline{\bf u}^{}_-\,)\ ,\ee
and

\be
m^{}_d(\,e^{-i\eta_d}_{}\,{\bf d}^{}_+\overline{\bf d}^{}_-+e^{i\eta_d}_{}{\bf d}^{}_-\overline{\bf d}^{}_+)+m^{}_d\delta^{}_d(\,{\bf d}^{}_+\overline{\bf d}^{}_++ {\bf d}^{}_-\overline{\bf d}^{}_-\,)\ .\ee
As a result of the perturbation ($\delta \ll 1$), the masses are shifted to

\be
m^{}_{u_1}~=~m^{}_u(\, e^{-\eta_u}-\frac{\delta_u^2}{2\sinh\eta_u}\,)\ ;\qquad  
m^{}_{u_2}~=~m^{}_u(\,e^{\eta_u}+\frac{\delta_u^2}{2\sinh\eta_u}\,)\ ,
\ee
and similarly for the down quarks with $u\rightarrow d$. The 
left-handed mixing matrices are simply

\be
\mathcal{U}^{}_{2/3}~=~\pmatrix{\cos\theta_{u}^{}&\sin\theta_{u}^{}\cr -\sin\theta_{u}^{}&\cos\theta_{u}^{}}\,\frac{1}{\sqrt{2}}\pmatrix{1& 
i\cr 1& 
-i}\ ,
\ee 
with 

\be
\sin\theta^{}_u~\approx~\frac{\delta^{}_u}{2\sinh\eta_u}\ ,\ee
while 

\be
\mathcal{U}^{}_{-1/3}~=~\pmatrix{\cos\theta_{d}^{}&\sin\theta_{d}^{}\cr -\sin\theta_{d}^{}&\cos\theta_{d}^{}}\,\frac{1}{\sqrt{2}}\pmatrix{1& 
i\cr 1& -i}\ ,\ee 
with 

\be
\sin\theta^{}_d~\approx~\frac{\delta^{}_d}{2\sinh\eta_d}\ .\ee
It follows that the CKM rotation angle is 

\be
\theta~=~\theta^{}_u-\theta_d^{}~\approx~ \frac{\delta^{}_u}{2\sinh\eta_u}-\frac{\delta^{}_d}{2\sinh\eta_d} \ .\ee
In the limit of strong hierarchies, $\eta_{u,d}>>1$, we can write the
mixing angle as

\be
\theta~\approx~{\delta^{}_u}\sqrt{\frac{m_{u_1}}{m_{u_2}}}-{\delta^{}_d}\sqrt{\frac{m_{d_1}}{m_{d_2}}} \ .\ee
Unfortunately, we cannot estimate its value since $\delta_{u,d}$ are arbitrary. 
 
These instructive two-family models demonstrate how large angles can arise
from a {\em slightly broken continuous family symmetry}.  With exact symmetry, the $\Delta I_w=1/2$ masses are hierarchical and the light neutrinos are mass-degenerate after the seesaw, even with degenerate $\Delta I_w=0$ masses.  The symmetry breaking schemes yield large lepton mixings, with small deviations from maximal mixing.  The approach lends itself to a further embedding in a grand unified picture, as the structure of the  $\Delta I_w=1/2$ mass matrices yields a good theoretical starting point for describing the hierarchical quark and charged lepton masses.

We note here that the idea of obtaining large angles using $U(1)$ family symmetries has been previously explored in the literature; an explicit example is the two-family scenario of \cite{Allanach:1998xi}.  However, our scenario has several important differences from this model and most other Abelian flavor models.  First, there is a clear difference in the philosophy of our approach, which is to motivate symmetries by requiring degeneracies, rather to use the usual flavor model-building methodology, which is to demand a family symmetry and explore its resulting implications.  This difference, though subtle, is one which does motivate the particular two-family $SO(2)$ scenarios presented here, which display a certain simplicity and elegance (e.g. with simple charge assignments).  The $SO(2)$ models also have an intriguing potential embedding in the braneworld context, given the hierarchical Yukawa couplings required at Froggatt-Nielsen tree level.\footnote{Note that higher-order FN effects or symmetry breaking dynamics are often required to achieve hierarchies in flavor models, but are not needed here (at least not at leading order).  A complete exploration of these effects is beyond the scope of this paper.} 
However, the most important difference is the requirement of {\it complex} Dirac mass matrices, which singles out a {\it nonstandard flavor basis with nontrivial phases} (the $\psi \pm  i\psi^\prime$ combination of fields as opposed to the standard $\psi$, $\psi^\prime$ basis). In most $U(1)$ flavor models, there is no natural explanation for these phases; they are just part of the usual $O(1)$ coefficients which Abelian symmetries do not fix.  For this reason, these phases, which play a  crucial role in our approach) are generically not found in \cite{Allanach:1998xi} or in other Abelian flavor models in the literature.

 We next discuss possible generalizations of this mechanism to three families.

\section{Extensions to Three Families}
\label{threefamsect}

For three families, the naive generalization of the $O(2)$ symmetry which naturally emerges in the two-family case to $O(3)$ does not allow for hierarchical Dirac mass matrices and degenerate neutrinos along the lines of our approach, as discussed in Section~\ref{mainapproach}.  The question of interest, therefore, is how to generalize the elegant mechanism of the two-family model based on $SO(2)$ to the case of three-family mixing.  The main theoretical challenge is to reproduce (or more optimistically, explain) the intriguing pattern of mixing angles seen in the data.

Although a full $O(3)$ symmetry does not satisfy the requirements of our theoretical approach,  there are other ways to generalize the approach to three family models.  One way is to keep the same $SO(2)$ symmetry and simply assign family charges to the third family fields in a way which incorporates the two-family mixing results. Another is to seek alternate symmetries which may or may not include the family charge of the previous section.

\subsection*{$SO(2)$ Models}
We begin with the $SO(2)$ models.  In this case, we simply assign $SO(2)$ charges to the fields of a
third family, $L^{\prime\prime}$, $\overline{e}^{\prime\prime}$,
$\overline{N}^{\prime\prime}$, $\bf{ Q}^{\prime\prime}$, $\bf{\overline
u}^{\prime\prime}$, and $\bf{\overline d}^{\prime\prime}$.   One possibility is that the third family fermions are uncharged under the $SO(2)$.  
If, on the other hand, the third family fields have nonvanishing $SO(2)$ charges, 
 each third family field $\psi^{\prime\prime}$ either has the same charge as $\psi_+$ or $\psi_-$ (and mix with these states in the symmetry limit), {\it i.e.}:
 $$\psi^{\prime\prime}\rightarrow e^{\pm i\alpha}\psi^{\prime\prime}.$$
In what follows, we will provide a brief summary of these two situations, relegating details and explicit models to a more comprehensive analysis given in \cite{pierrelisalong}. \\

\noindent $\bullet$ {\bf Vanishing third family $SO(2)$ charges.}\\
When the third family is blind to the $SO(2)$ symmetry, we will see that the path to degeneracy is arguably the most straightforward.  For the neutral leptons,  it is straightforward to see that degenerate right-handed neutrinos arise in the limit of exact family symmetry when $\overline{N}^{\prime\prime}$ and  $L^{\prime\prime}$ are uncharged under the $SO(2)$:
 \be
2 M \overline{N}_+\overline{N}_-+M_0\overline{N}^{\prime\prime}\overline{N}^{\prime\prime}.
\ee
$M_0=M$ corresponds to $SO(3)$ invariance in the electroweak singlet sector, which could arise from a custodial symmetry which is present in this sector but not in  
the electroweak doublet sector.  The neutral Dirac couplings are 
\be
\mu e^\eta \nu_+\overline{N}_-+\mu e^{-\eta}\nu_-\overline{N}_+ +\mu_0\nu^{\prime\prime}\overline{N}^{\prime\prime}.
 \ee
The Dirac mass eigenvalues are nonvanishing and hierarchical.  In the symmetry limit, all three neutrinos seesaw:   
$$m_{1,2}=\frac{\mu^2}{M},\;\;\;\;\;m_3=\frac{\mu_0^2}{M_0}.$$
Two of the masses are naturally degenerate, but total degeneracy can be achieved depending on the parameters of the theory.  The mixing angles and detailed mass splittings then depend on the form of the symmetry breaking,  which can occur in the $\Delta I_w=0$ and/or $\Delta I_w=1/2$ sectors. For example, in the right-handed neutrino sector one can have symmetry breaking terms which couple $\overline{N}^{\prime\prime}$ to $\overline{N}_\pm$ ({\it i.e.} $\Delta F=\pm 1$ effects, in which $F$ denotes the family charge), or corrections to the couplings of the $\overline{N}_\pm$ sector ($\Delta F=2$ effects).  The main theoretical challenge in both cases is to achieve the two large mixing angles while keeping $\theta_{13}$ small, which constrains the allowed symmetry breaking effects.\\

\noindent $\bullet$ {\bf Nonvanishing third family $SO(2)$ charges.} \\ In this case, the third family fields are charged under the $SO(2)$ and mix with the other families in the symmetry limit.  For reasons which will become clear shortly, 
we parametrize this mixing as
\bea
\psi_\pm(\zeta,\phi)&=&\cos\zeta \psi_\pm +\sin\zeta e^{i\phi}\psi^{\prime\prime}\nonumber \\
\psi_\pm^\perp(\zeta,\phi)&=&-\sin\zeta e^{-i\phi} \psi_\pm +\cos\zeta\psi^{\prime\prime},
\eea
in which $\zeta$ and $\phi$ are arbitrary angles.  
Since certain Standard Model fields enter more than one type of mass term (e.g. lepton doublets enter both the charged and neutral Dirac mass terms, and right-handed neutrinos enter both the neutral Dirac and Majorana masses), when the third family leptons have $SO(2)$ charges, the phenomenological discussion can be cast into the question of which linear combinations of fields enter the different types of mass terms.

When the {\it same} linear combinations enter two types of Yukawa couplings, we say that they are ``locked."    A familiar example of near locking is that of the quark doublets, which enter the up-type and down-type Yukawa couplings in nearly the same way up to Cabibbo-sized effects; {\it i.e.}, the CKM matrix measures the deviation from perfect locking.  For the lepton sector, if the seesaw mechanism is operational the observed large angles of the MNSP matrix do not preclude the possibility of locking in either the $L$ or $\overline{N}$ sectors. 

Locking can be achieved by invoking a second family phase symmetry $U(1)_\beta$,  under which e.g. for the $\overline{N}$ sector (assuming for concreteness that $\overline{N}^{\prime\prime}$ and $\overline{N}_-$ have the same charges under the original $SO(2)_\alpha$): 
$$(\overline N^{}_--i\overline{N}^{\prime\prime})~\rightarrow~~e^{-i\,\beta}_{}\,(\overline{N}_--i\overline
N^{\prime\prime})\ ,\qquad (\overline N^{}_-+i\overline
N^{\prime\prime})~\rightarrow~~e^{i\,\beta}_{}\,(\overline 
N^{}_-+i\overline
N^{\prime\prime}),$$
which requires the same linear combination of $\overline{N}$'s to enter the  $\Delta I_w=0$ and the $\Delta I_w=1/2$ couplings.
Similar considerations apply to the lepton doublets.

With this in mind, note that there are two coupling schemes of interest, depending on the relative signs of the charges of $L^{\prime\prime}$ and $\overline{N}^{\prime\prime}$:\\

\noindent $\bullet$ {\it Type A couplings}.  In this coupling scheme, $L^{\prime\prime}$ and $\overline{N}^{\prime\prime}$ have the same nonvanishing $SO(2)$ charge; we assume for concreteness that they mix with $L_-$ and $\overline{N}_-$, respectively.  The $SO(2)$ invariant $\Delta I_w=0$ mass terms are then
\be 
\label{typeAmaj}
M \overline{N}_+\overline{N}_-(\zeta,\phi).
\ee
As only one combination of $\overline{N}_-$ and $\overline{N}^{\prime\prime}$ enters in the Majorana mass terms, the orthogonal combination $\overline{N}_-^\perp(\zeta,\phi)$ does not participate in the seesaw.  Similarly, it is straightforward to see that without loss of generality, the $SO(2)$ invariant neutral $\Delta I_w=1/2$ mass terms are of the form
\be
\mu e^\eta  \nu_+\overline{N}_-(\zeta',\phi')+\mu e^{-\eta} \nu_-(\chi,\xi)\overline{N}_+.
\ee
As only one combination of $\overline{N}$'s and $L$'s enters the Dirac mass terms, $\nu^\perp(\chi,\xi)$ and $\overline{N}^\perp(\zeta,\phi)$ remain massless in the $SO(2)$ symmetry limit.   The physics also depends on whether there is locking.  If locking takes place in the $\overline{N}$ sector, $\zeta=\zeta^\prime$ and $\phi=\phi^\prime$.  Alternatively, there could be locking in the $L$ sector, such that the combination $e_-(\chi',\xi')$ which enters the charged lepton couplings (which depends on the charges of the $\overline{e}$'s; there is always an orthogonal combination $e_-^\perp(\chi',\xi' )$ which does not enter) satisfies the conditions $\chi=\chi^\prime$ and $\xi=\xi^\prime$. 

With or without locking,  the pattern of masses and mixings in the limit of exact $SO(2)$ symmetry is unrealistic, but this can be remedied by symmetry breaking.   If symmetry breaking occurs only in the $\Delta I_w=0$ sector, the absence of $\nu^\perp$ in the Yukawa couplings implies one massless 
left-handed neutrino: small perturbations (minimally to give $\overline{N}^\perp$ a large Majorana mass) will produce the so-called inverted hierarchy. In this case, the neutrino mass splitting is related to the haze away from maximal mixing, and applies to solar oscillations. The haze turns out to be too small to explain the solar angle, although it naturally yields $\theta_{13}=0$ in the symmetry limit.   The atmospheric angle is not determined, but instead can be directly traced to the absence of locking in the $L$ sector ({\it i.e.}, $\theta_\oplus \sim \chi^\prime -\chi$).  While these minimal models are not consistent with the data, more promising models can be constructed by including $\Delta I_w=1/2$ symmetry breaking couplings that bring $\nu^\perp$ into the neutrino seesaw.\\

\noindent $\bullet$ {\it Type B couplings.}  In this class of models, $L^{\prime\prime}$ and $\overline{N}^{\prime\prime}$ have nonvanishing but opposite $SO(2)$ charges; e.g., let us assume for concreteness that $\overline{N}^{\prime\prime}$ mixes with $\overline{N}_-$ and $L^{\prime\prime}$ mixes with $L_+$.    In this case, the $\Delta I_w=0$ mass terms are given by Eq.~(\ref{typeAmaj}) as before.  However, the $\Delta I_w=1/2$ couplings are different:
\be
\mu e^\eta\,\nu^{}_+(\chi,\xi)\,{\overline N}^{}_-(\zeta',\phi') 
~+~\mu e^{-\eta}\,\nu^{}_-\,{\overline N}^{}_+ 
+\,\mu^\prime\,\nu_+^\perp(\chi,\xi)\,\overline 
N_-^\perp(\zeta',\phi').
\ee
In the above, $\mu^\prime$ is assumed to be of the order of the electroweak scale ({\it i.e.}, $\mu^\prime$ is of $\mathcal{O}(\mu)$, as motivated by the logarithmic spacing of the quark and charged lepton masses), but otherwise is a free parameter. Since both $\nu^\perp$ and $\overline{N}^\perp$ have Yukawa couplings, the neutral Dirac matrix has no zero eigenvalues. Furthermore, there are no massless neutrinos in the $SO(2)$ symmetry limit. Two of the neutrinos have seesaw suppressed masses, while the other two neutrinos generically have electroweak scale masses, whether or not there is $\overline{N}$ locking.   

The neutrino masses and mixings are again not realistic in the symmetry limit for Type B couplings.  With $\Delta I_w=0$ symmetry breaking terms (again minimally to give $\overline{N}^\perp$ a large Majorana mass), the phenomenological implications of the resulting models depend strongly on whether $\overline{N}$ locking occurs or not.  With $\overline{N}$ locking,  $\theta_{13}$ is naturally zero (which generically occurred for Type A models) independently of the form of the symmetry breaking terms in the $\Delta I_w=0$ sector. However, the Type B models also do not require an inverted hierarchy scheme, and hence lead to a greater flexibility for constructing viable models.  For example, models can be constructed with minimal tuning in which the two-family mechanism described in Section~\ref{twofamsect} can be successfully applied to atmospheric oscillations; the solar angle is then directly related to the absence of locking in the $L$ sector.  Without $\overline{N}$ locking,  there is again a greater flexibility to construct models due to the various possibilities for the mass hierarchy.  However,  in this case generically $\theta_{13}\neq 0$, and hence the theoretical challenge is to obtain the two large mixings while keeping $\theta_{13}$ small.

For both Type A and B couplings,  only two right-handed neutrinos are degenerate; {\it i.e.},  $\mathcal{M}_{\rm Maj}\neq 1$ in the symmetry limit.  Hence, when the third family fields are charged under the original $SO(2)$ symmetry, the resulting models step outside of our theoretical framework.  For Type A models, light neutrino mass degeneracy will not occur unless there is a conspiracy between specific symmetry breaking terms in both the $\Delta I_w=0$ mass terms (to give the third right-handed neutrino a mass) and $\Delta I_w=1/2$ couplings (to involve the third family in the seesaw).  For Type B couplings, degeneracy of the light neutrinos requires symmetry breaking in the $\Delta I_w=0$ sector and a specific tuning of the parameters. \\

In summary, the three-family $SO(2)$ schemes provide an interesting framework for model building, but have the negative generic feature that neutrino mass degeneracy is lost (for both light and heavy neutrinos).  The lack of degeneracy is also a hallmark of more general attempts to obtain viable three-family mixing schemes with Abelian family symmetries.  Despite this shortcoming, our approach may provide new insights or arise in particular settings.  Further exploration of these scenarios will be presented in \cite{pierrelisalong}.

\subsection*{Outlook: The Path to Neutrino Mass Degeneracy}
We have shown in this paper that for two families, it is possible to obtain neutrino mass degeneracy (via the seesaw) through a continuous family symmetry, while allowing 
for hierarchical charged fermion masses.  However, as discussed in Section~\ref{mainapproach}, generalization to three families has proven more challenging.  Let us recall the constraints imposed on the
Dirac and Majorana matrices by requiring for simplicity that the seesaw combination
\be
M^{}_{D}\frac{1}{\cal M_{\rm Maj}}\,M^T_D 
\ee
is proportional to the identity matrix (a stronger condition than was previously imposed in Section~\ref{mainapproach}). If we write the Dirac matrix in terms
of three vectors ${\bf v}_{[i]}$, we can think of the inverse Majorana
matrix as a metric in this three dimensional space. In its ``flat" limit,
the vectors are orthogonal

\be
 {\bf v}^T_{[i]}\,{\bf v}^{}_{[j]}~\sim~\delta^{}_{ij}\ .\ee
If the vector entries were all real, the Dirac matrix would simply be a 
rotation matrix, in which case its three eigenvalues would be equal,
and would not produce any hierarchy. In the $(2\times 2)$
case, we could produce a hierarchy by introducing a family $SO(2)$, but
then the vectors had complex entries.

In the $(3\times 3)$ case, a matrix that fulfills our requirement is 

\be
{\cal M}^{}_D~=~\pmatrix{\cosh\eta&i\sinh\eta\cos\theta&i\sinh\eta\sin\theta\cr
-i\sinh\eta&\cosh\eta\cos\theta&\cosh\eta\sin\theta\cr 
0&-\sin\theta&\cos\theta}\ .\ee
Its eigenvalues ($1\ ,~e^{\,\eta}\ ,~e^{-\eta}$) show equal logarithmic
spacing of the masses, clearly a desirable feature for charged fermion Dirac matrices. Unfortunately, we have not found a symmetry principle from which this naturally emerges.

A second approach is to invoke non-Abelian family symmetries. 
For three families, the splitting of the neutrino masses from full degeneracy can be viewed as a small $\lambda_8$ breaking and smaller $\lambda_3$ breaking, suggesting an $SU(3)$ family symmetry, or 
a small breaking along $m^2$ and smaller breaking along $m$, where $m$ is the flavor magnetic quantum number of an $SO(3)$ family symmetry.

We have already discussed the shortcomings of a full family $SO(3)$ here.  A family $SU(3)$ can naturally describe the splittings, but the path to degeneracy is again quite different than our approach. With one right-handed neutrino per family, the Majorana mass matrix transforms
as a family sextet and can be highly hierarchical (see e.g. \cite{flavorsu3,khlopov}). 
However, we can augment the number of $\overline{N}$'s: for example, $E_6$ requires two right-handed neutrinos per family.

In such models, an amusing possibility is to invoke an additional phase symmetry which 
allows the six neutrinos to be arranged into one triplet and one
antitriplet. Invariance under this phase symmetry forces their mass
matrix to be either a family singlet or a family octet. In the limit of
exact 
symmetry, the three left-handed neutrinos stay massless,
achieving degeneracy. After symmetry breaking, they acquire masses through the seesaw. The
light neutrino mass splittings can then be described by perturbations along $\lambda_8$ and $\lambda_3$. Although this deviates from the specific theoretical approach of this paper, it is an intriguing path to degeneracy which may provide a fruitful ground for model building.
We hope to return to these issues elsewhere.\\

In summary, we have outlined a flavor model building framework in which continuous family symmetries emerge from the theoretical requirements of hierarchical charged fermion masses and degenerate neutrino masses for both heavy and light neutrinos. Neutrino mass degeneracy is the starting point because large mixings can arise from small perturbations about degenerate structures, and quark-lepton unification suggests that such small parameters should be expected in the lepton sector as well as in the quark sector.

The requirements of degeneracy with hierarchy naturally stem from symmetry for two families, but for 
 three families this does not naturally emerge and finding a symmetry-based starting point is more difficult.  However, one should keep in mind that it is a formidable challenge to
explain the three-family lepton mixing pattern together with the quarks in a truly satisfactory way, which is why the flavor puzzle has taken such an intriguing turn in light of the recent lepton data. Although there are challenges in extending our
approach to three families which mirror those found in general $U(1)$
models, the approach is incremental, and it does show what we must do to
achieve naturalness in a three family context.  The work therefore provides a
particular setting in which to investigate these issues which may yield insight or naturally occur in specific theoretical contexts, and as such is one of many possible avenues worthy of further exploration in addressing the flavor puzzle of the Standard Model.


\section*{Acknowledgments} 
We thank V. Barger, D. Chung, D. Demir, G. Ross, and C. Thorn for helpful discussions.   P.R. thanks the Aspen Center for Physics and L.E. thanks the University of Wisconsin-Madison for hospitality during the completion of this work.  L.E. also thanks the L'Or\'{e}al for Women in Science U.S. Postdoctoral Fellowship Program for its generous support.  This work is supported by the U.S. Department of Energy under the grant DE-FG02-97ER41209.

\end{document}